\documentclass[article]{aa}

\usepackage{natbib}
\usepackage{graphicx}
\usepackage{txfonts}

\begin{document}

\title{Transverse kink oscillations in the presence of twist}
\author{J. Terradas$^{1}$, M. Goossens$^{2}$ }

\offprints{J. Terradas, \email{jaume.terradas@uib.es}}
\institute{$^1$ Departament de F\'\i sica,
Universitat de les Illes Balears, E-07122, Spain, \email{jaume.terradas@uib.es} \\
$^2 $Centre for Plasma Astrophysics,  Department of Mathematics, Celestijnenlaan
200B,  B-3001, Katholieke Universiteit Leuven, Leuven, Belgium, \email{Marcel.Goossens@wis.kuleuven.be}}
{}

\date{Received / Accepted }

\abstract{Magnetic twist is thought to play an important role in coronal loops.
The effects of magnetic twist on stable magnetohydrodynamic (MHD) waves is
poorly understood because they are seldom studied for relevant cases.} {The goal
of this work is to study the fingerprints of magnetic twist on stable transverse
kink oscillations.} {We numerically calculated the eigenmodes of propagating and
standing MHD waves for a model of a loop with magnetic twist. The azimuthal
component of the magnetic field was assumed to be small in comparison to the
longitudinal component. We did not consider resonantly damped modes or kink
instabilities in our analysis.} {For a nonconstant twist the frequencies of the
MHD wave modes are split, which has important consequences for standing waves. 
This is different from the degenerated situation for equilibrium models with
constant twist, which are characterised by an azimuthal component of the
magnetic field that linearly increases with the radial coordinate.}{In the
presence of twist standing kink solutions are characterised by a change in
polarisation of the transverse displacement along the tube. For weak twist, and
in the thin tube approximation, the frequency of standing modes is unaltered and
the tube oscillates at the kink speed of the corresponding straight tube. The
change in polarisation is linearly proportional to the degree of twist. This
has implications with regard to observations of kink modes, since the detection
of this variation in polarisation can be used as an indirect method to estimate
the twist in oscillating loops.}


\titlerunning{Transverse MHD waves in the presence of twist}
\authorrunning{Terradas et al.}
\maketitle
\newcommand{\etal}{{et al.}}

\section{Introduction} 

It is anticipated that magnetic twist plays a considerable role in the structure
of coronal loops. The highly dynamic photosphere and chromosphere can introduce
twist in the magnetic field of the solar corona. For example, footpoint rotation
and shear motion may result in twisting of coronal loops \citep[see
e.g.,][]{brownetal2003}. A newly emerged magnetic field is also supposed to be
twisted during the buoyant evolution through the convection zone, meaning that
loops may emerge already twisted \citep[see
e.g.,][]{morenoemonet96,hoodetal2009}. Twist also considerably contributes to
the eruption of many prominences and CMEs (coronal mass ejections) \citep[see
e.g.,][]{hoodpriest79,priestetal89}.

In the theoretical analysis of magnetohydrodynamic (MHD) waves, the effects of a
curved loop axis are usually omitted. The stability of straight  tubes with
twisted magnetic field has been investigated by e.g.,  \citet{shafranov57},
\citet{kruskaletal58}, and \citet{suydam58}. The role of line tying conditions
in the context of coronal loops has been studied by e.g., \citet{raadu72} while
the effect of gas pressure on the kink instability has been analysed by e.g.,
\citet{giachettietal77}. The combination of both effects has been studied by
\citet{hoodpriest79}, \citet{hoodpriest81}, \citet{ hoodetal82},
\citet{einaudivanhoven83}, and \citet{vellietal90}. More recently,
\citet{temurietal2010} and \citet{diazetal11} have considered the stability of
twisted magnetic flux tubes with mass flows along the field lines. The main aim
of the previous works is determining the instability threshold and calculating
the growth rate of kink unstable modes.

Stable propagating MHD waves have been studied previously, but almost always
under particular conditions that somehow  reduce the mathematical complexity.
Often the MHD waves are incompressible and/or the azimuthal component of the
magnetic field varies linearly with distance up to a certain position where it
drops to zero or all the way up to infinity. The first choice leads to a surface
current while the second choice causes magnetic energy to diverge. Early studies
were performed by \citet{goossensetal92} and \citet{bennettetal99} in the
incompressible regime. The analysis of sausage modes ($m=0$) has been carried
out by \citet{robfedun06,robfedun07}. Kink modes assuming compressible motions
have been considered by \cite{robcart06,cartrob08,robfedun10}. In these works
standing transverse waves have not been considered and only propagating waves
have been analysed.

In particular, \cite{ruderman07}  \citep[see also review by][]{ruderrob09}
studied standing waves and concluded that twist does not affect kink modes,
only fluting modes are modified \citep[see also][]{goossensetal92}. This result
is due to the particular choice of constant magnetic twist, i.e., twist is
independent of the radial coordinate. The conclusions of the work of
\cite{ruderman07} have some authors to regard twist as unimportant
for transverse kink oscillations. 

In this paper we study the effect of an azimuthal magnetic field component on
the frequencies of MHD waves for an equilibrium model with nonconstant magnetic
twist. We are especially interested in the effect of twist on standing kink
modes, a problem that, to our knowledge, has not been addressed in the context
of transverse coronal oscillations \citep[with the exception of standing fluting
modes, studied in][]{ruderman07}. Thus, our focus is on the fingerprints of
twist on transverse kink oscillations but in the stable regime. This might be
relevant for interpreting  kink oscillations, from the early transverse kink
oscillations reported with TRACE \citep{aschetal99,nakaetal99}, to the last AIA
observations of transverse oscillating loops \citep{aschwetal2011}.


\section{Equilibrium model of the magnetically twisted loop}\label{equilsec}

The model for the coronal loop that we use in the present paper is a straight
cylinder with constant density inside the cylinder and constant density outside
the cylinder. Thus, the coronal loop is modelled as a density enhancement characterised by a density
contrast $\rho_{\rm i}/\rho_{\rm e}$ with $\rho_{\rm i}$ being the internal density  and $\rho_{\rm e}$
the external density, 
\begin{eqnarray}\label{dens}
\rho=\left\{
\begin{array}{ll}
\rho_{\rm i}, & \; 0\leq r< R, \\
\rho_{\rm e}, & \; r>R.
\end{array}
\right.
\end{eqnarray}
At $r=R$ the density changes in a discontinuous manner and $R$ is the radius of
the loop.

Under coronal conditions it is a good approximation to neglect plasma pressure
compared to magnetic pressure. This zero-$\beta$ approximation removes the slow
magnetoacoustic waves from the analysis. In the present paper we allow the magnetic twist to be nonconstant. The
model starts from the equation of magnetostatic equilibrium
\begin{eqnarray}
\frac{d B^2}{dr}=-2\frac{B_\varphi^2}{r},
\label{eqmagn}
\end{eqnarray}
where
\begin{eqnarray}
B^2=B_z^2+B_\varphi^2.
\label{eqbtot}
\end{eqnarray}
Here the azimuthal component of the magnetic field is prescribed and
Eq.~(\ref{eqmagn}) is solved for $B_z(x)$ with $x=r/R$. 
The solution is 
\begin{eqnarray}
B_z^2(x)=B_z^2(0)-B_\varphi^2(x)-2 \int_0^x \frac{B_\varphi^2(s)}{s} ds.
\label{eqbz}
\end{eqnarray}
The azimuthal component of the field is written as
\begin{eqnarray}
B_\varphi(x)=A f(x),
\label{eqbphi}
\end{eqnarray}
where $A$ has the dimension of a magnetic field and $f(x)$ is a dimensionless
function satisfying the condition $f(0)=0$ or better $f(x)\rightarrow 0$ for   
$x \rightarrow 0$. If we insert Eq.~(\ref{eqbphi}) in Eq.~(\ref{eqbz}) we find
\begin{eqnarray}\label{eqbzsol}
B_z^2(x)&=&B_z^2(0)-A^2 G(x),\\
G(x)&=&f^2(x) + 2 \int_0^x \frac{f^2(s)}{s} ds.
\label{eqgsol}
\end{eqnarray}
Notice that $G(0)=0$ and $G(x) \ge 0$.

A particular choice for $f(x)$ used in the present paper is
\begin{eqnarray}
f(x)=\left\{
\begin{array}{lll}
0, & \; 0\leq x< p, \\
\left(x-p\right)\left(q-x\right), & \; p\leq x \leq q, \\
0, & \; x>q,
\end{array}
\right.\label{eqffunct}
\end{eqnarray}
where $p$ and $q$ are constants that we can freely choose. The choice
$p=0$ is an option and that $q>1$ is also allowed. For this model we have the
freedom to have twist inside and also outside the tube. The function $f(x)$ attains its maximal value $f_{max}=(q-p)^2/4$ at $x_m=(p+q)/2$.
Hence
\begin{eqnarray}
B_{\varphi}(x)=\alpha B_z(0) \frac{f(x)}{f_{max}}.
\label{eqBphimax}
\end{eqnarray}
In order to relate the strength of the azimuthal component $B_\varphi$ with
respect to the dominant longitudinal component, we introduce $\alpha$:
\begin{eqnarray}
\alpha=\frac{B_{\varphi, max}}{B_z(0)},
\label{eqalph}
\end{eqnarray}
hence 
\begin{eqnarray}
A=\frac{\alpha}{f_{max}} B_z(0).
\label{eqA}
\end{eqnarray}
For the function $f(x)$ specified by Eq.~(\ref{eqffunct}) it follows that
\begin{eqnarray}
G(x)=\left\{
\begin{array}{lll}
0, & \; 0\leq x< p, \\
>0, & \; p\leq x \leq q, \\
G(q), & \; x>q.
\end{array}
\right.
\end{eqnarray}
This function is calculated using Eq.~(\ref{eqgsol}).

In summary, the equilibrium magnetic field is defined as
\begin{eqnarray}
B_z=B_0, \,\, B_\varphi=0,\label{b0}
\end{eqnarray}
for $x<p$,
\begin{eqnarray}
B_z=\sqrt{B_0^2-A^2 G(x)}, \,\, B_\varphi=A (x-p) (x-q),
\end{eqnarray}
for $p<x<q$, with  
 \begin{eqnarray}
G(x)=(x-p)^2 (x-q)^2+c_4(x^4-p^4)+c_3 (x^3-p^3)\nonumber \\
+c_2 (x^2-p^2)+c_1(x-p)+c_l \left(\ln(x)-\ln(p)\right),
\end{eqnarray}
\begin{eqnarray}
c_1&=&-4pq(p+q),\\
c_2&=&(p+q)^2+2 p q,\\
c_3&=&-\frac{4}{3}(p+q),\\
c_4&=&\frac{1}{2},\\
c_l&=&2 p^2 q^2,
\end{eqnarray}
and 
\begin{eqnarray}
B_\varphi=0, \,\, B_z=\sqrt{B_0^2-A^2 G(q)},
\end{eqnarray}
for $x>q$.

\section{MHD Eigenmodes}\label{secteigenmodes}

\subsection{MHD equations and numerical method}

The  motions superimposed on the equilibrium model given in the previous section
are governed by the linearised ideal MHD equations in cylindrical coordinates. 
These equations can be found in  the appendix. Several terms involving radial
derivatives of the equilibrium magnetic field are present in the equations. 

In the MHD equations we have performed a Fourier analysis in time, with $\omega
$ the frequency, and also in the azimuthal direction, $m$ being the azimuthal
wavenumber. For the longitudinal direction (along the tube axis) two situations
are studied. The first situation corresponds to fully propagating waves,
allowing Fourier decomposition in $z$, $k$ being the longitudinal wavenumber. In
this case the equations are solved in the radial coordinate $r$. The second
situation corresponds to standing waves with line-tying conditions at the
footpoints of the loop. This problem does not allow Fourier analysis in $z$, and
the equations are solved in 2D (in $r$ and $z$). For the analysis of propagating
waves the system of equations reduces in the ideal regime to the Hain-L\"ust
equation. In general, it is not possible to find analytical solutions to this
equation when the twist profile is nonconstant. For this reason the MHD
equations are numerically solved using the code PDE2D \citep{sewell2005} in the
two situations (1D and 2D) explained above. More details about the numerical
method can be found in \citet{terretal06b}. In the present paper we avoided 
possible Alfv\'enic resonances on purpose, interesting as they might be. In
addition, we stayed away from a strong azimuthal field that can cause kink
instabilities of the magnetic flux tube. This means that the eigenfrequency
$\omega$ is a purely real magnitude in our analysis.

\subsection{Effect of the azimuthal magnetic field on MHD
waves}\label{secteigenmodestwist}

Since we adopted the zero-$\beta$ approximation, the only characteristic frequency in
the system is the local Alfv\'en frequency $\omega_{\rm A}$. The effect that the
azimuthal component of the magnetic field $B_\varphi$ has on the MHD waves
can be appreciated from the effect of $B_\varphi$ on $\omega_{\rm A}^2$. The
square of the Alfv\'en frequency is defined as
\begin{eqnarray}
\omega_{\rm A}^2=\frac{1}{\mu_0 \rho} \left(\frac{m}{r} B_\varphi+ k
B_z\right)^2.
\end{eqnarray}
This expression can be written as 
\begin{eqnarray}\label{eqalfvtwist}
\omega_{\rm A}^2=\frac{B_z^2}{\mu_0 \rho}
\frac{1}{R^2}(kR)^2 \left(1+\frac{m}{\pi}\Phi(r)\right)^2.
\end{eqnarray}
The function $\Phi(r)$ is 
\begin{eqnarray}
\Phi(r)=\frac{\pi}{k R}\frac{B_\varphi}{(r/R) B_z}=L\frac{B_\varphi}{r B_z},
\end{eqnarray}
where we have defined $L$ that $k=\pi/L$. $\Phi(r)$ is the twist of the magnetic
field lines and it is $2\pi$ times the number of windings of the field around
the loop axis over a distance $L$. Equation~(\ref{eqalfvtwist}) suggests that even a relatively
small $\varphi-$component of the magnetic field in the sense that
$|B_{\varphi}|<<|B_z|$, might have a relatively important effect on the MHD
waves. The relative importance of $B_{\varphi}$ to $B_z$ is
controlled by the amount of twist of the magnetic field, with the length $L$
being a relevant factor.

The ratio of the contribution to the Alfv\'en frequency due to the azimuthal
component $B_\varphi$ to that due to the longitudinal component is
\begin{eqnarray}\label{ratio}
\frac{m}{k R}\frac{B_\varphi}{(r/R) B_z}.
\end{eqnarray}
We computed the maximal value of the ratio for the model discussed in
Section~\ref{equilsec}. We considered a weak
azimuthal field and a relatively low value of $\alpha<<1$. To first order in
$\alpha$ (small $\alpha$ approximation) 
\begin{eqnarray}
B_z&\simeq& B_z(0),\\
B_\varphi&=&\alpha B_z(0) f(x)/f_{max}.
\end{eqnarray}
Hence
\begin{eqnarray}
\frac{B_\varphi}{(r/R) B_z}=\alpha \frac{f(x)}{x}\frac{1}{f_{max}}.
\end{eqnarray}
For the prescription of $f(x)$ given by Eq.~(\ref{eqffunct}) we already know the value
of $x_m$ and $f_{max}$. In addition, the function $f(x)/x$ is maximal at
$x_*=\sqrt{pq}$, and $f(x_*)/x_*=p+q-2\sqrt{p q}$. Hence the maximal value 
of 
\begin{eqnarray}
\left.\frac{B_\varphi}{(r/R) B_z}\right|_{max}=\alpha
\frac{4}{\left(q-p\right)^2}\left(p+q-2\sqrt{p q}\right).
\end{eqnarray}
For example, for the values $p=1/2$ and $q=3/2$ used in our calculations
we find that 
\begin{eqnarray}
\left.\frac{m}{k R} \frac{B_\varphi}{(r/R) B_z}\right|_{max}\approx 1.08\, m
\,\frac{\alpha}{kR}.
\end{eqnarray}
As an example we consider a loop that is 10 times longer than the radius. For the
fundamental mode along the tube axis it follows that $kR=\pi/10$. So for $\alpha=1/10$
the ratio is about 0.34. Thus, although the azimuthal component of the magnetic
field is weak, its effect on the Alfv\'en frequency is substantial. For
realistic loops it is a competition between the two quantities $kR$ and
$\alpha$ which determine the ultimate effect.

We omitted resonantly damped eigenmodes. This choice restricts the maximal twist
allowed in the system or conversely the allowed combinations of $\alpha$ and $k
R$. Even for a constant density profile like the one considered in this work
(see Eq.~(\ref{dens})), $\omega_{\rm A}$ is a function of position because
$B_\varphi$ varies with position and the corresponding term is multiplied by the
factor $1/r$. A necessary (but not sufficient) condition for the absence of
resonances is that $\max \omega_{\rm Ai}^2 \le \min \omega_{\rm Ae}^2$. This
conditions is equivalent to  \begin{eqnarray}\label{condnores} \max \Phi < \pi
\left(\sqrt{\rho_{\rm i}/\rho_{\rm e}}-1\right). \end{eqnarray} For $\rho_{\rm
i}/\rho_{\rm e}=3$ this implies that $\max \Phi \lesssim 0.73 \pi$.

Condition~(\ref{condnores}) can be rewritten in terms of the function $f(x)$ as
\begin{eqnarray}\label{eqalphaokR} \frac{\alpha}{k R}<x_*
\frac{f_{max}}{f(x_*)}\left(\sqrt{\rho_{\rm i}/\rho_{\rm e}}-1\right). \end{eqnarray}
For the typical values that we use in the computations the inequality is
$\alpha/{k R}\lesssim 0.67$. So we can define the critical value of $\alpha$ as
\begin{eqnarray}\label{eqalphaJ} \alpha_J=k R\, x_*
\frac{f_{max}}{f(x_*)}\left(\sqrt{\rho_{\rm i}/\rho_{\rm e}}-1\right).
\end{eqnarray}
Then for values $\alpha>\alpha_J$ the azimuthal component of the magnetic field
introduces resonances even in configurations with constant density. 
In addition to $\alpha_J$ we can consider the
situation when the azimuthal component $B_\varphi$ dominates the variation of 
$\omega_{\rm A}^2$. For $m=1$ this happens when $\max \Phi(r)\ge \pi$ (see
Eq.~(\ref{ratio})). For the
function $f(x)$ defined in Eq.~(\ref{eqffunct}) this can be reformulated into 
\begin{eqnarray}\label{eqalphaokRm} 
\frac{\alpha}{k R}\ge x_*
\frac{f_{max}}{f(x_*)}.
\end{eqnarray}
This forces us to define a second critical value for $\alpha$
\begin{eqnarray}\label{eqalphaM}
\alpha_M=k R\, x_*
\frac{f_{max}}{f(x_*)},
\end{eqnarray}
which differers from $\alpha_J$ by the factor $\left(\sqrt{\rho_{\rm
i}/\rho_{\rm e}}-1\right)$. For the present model with $p=1/2$ and $q=3/2$,
$\alpha_M \simeq 1.09 \, kR$.  Below
we used sufficiently low values of $\alpha$ compared to
$\alpha_J$ and $\alpha_M$. This implies that even in the thin tube approximation we can use the
small $\alpha$ approximation.

\section{Results: propagating waves}\label{sectprop}

\begin{figure}[!ht] \center{\includegraphics[width=8.5cm]{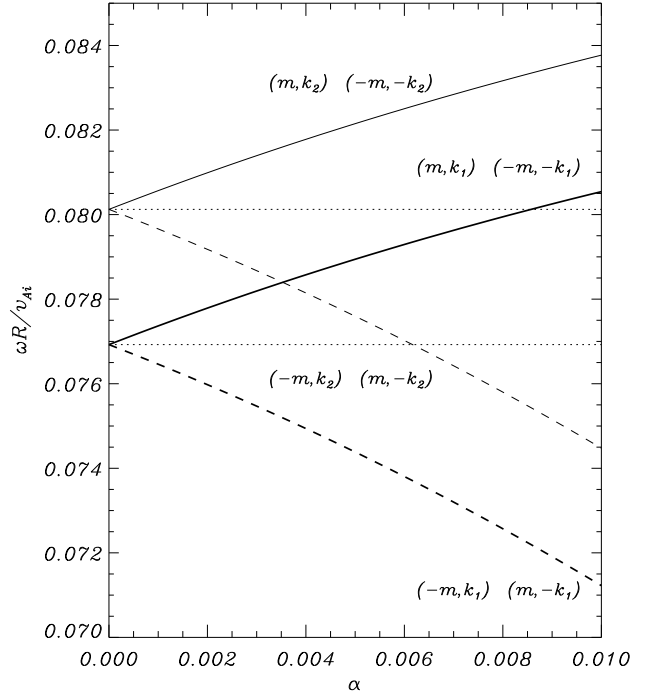}} 
\caption{\small Frequency as a function of twist ($\alpha$). The curves
are obtained by solving the eigenvalue problem in 1D. The dotted line represents
the solution for the untwisted case. For this particular example, $\rho_{\rm
i}/\rho_{\rm e}=3$, $p=1/2$, $q=3/2$. The two sets of solutions are associated
to $m=1$ and $k_1=\pi/50R$ (thick lines), and $k_2=\pi/48R$ (thin
lines).}\label{omegalamb} \end{figure}

We start by solving the eigenvalue problem for propagating waves using different
pairs of the wavenumbers $m$ and $k$, which are assumed to be positive. We use a
particular choice of parameters,  $\rho_{\rm i}/\rho_{\rm e}=3$, $p=1/2$,
$q=3/2$. In Figure~\ref{omegalamb} the dependence of the eigenfrequency on
$\alpha$, which measures the amount of twist, is plotted. For the pair $(m,k_1)$
the frequency is $\omega=k_1 c_{\rm k}$ when $\alpha=0$, i.e., we recover the
known result in the absence of twist (in the thin tube approximation), which is
the kink speed
\begin{eqnarray}\label{kinkspeed}
c_{\rm k}=\sqrt{\frac{2}{1+\rho_{\rm e}/\rho_{\rm i}}}v_{\rm Ai},
\end{eqnarray}
where $v_{\rm Ai}=B_0/\sqrt{\mu_0 \rho_{\rm i}}$ is the internal Alfv\'en speed
for a straight field.

Figure~\ref{omegalamb} shows that the frequency increases monotonically with
$\alpha$ (see continuous thick line). Exactly the same curve is obtained for the
wavenumbers $(-m,-k_1)$, meaning that the results are degenerate with respect to
a change in sign of the two wavenumbers. In constrast, for $(-m,k_1)$ and
$(m,-k_1)$ the frequency decreases with $\alpha$ (see dashed thick  line). This
behaviour reflects the fact that, in contrast to the purely vertical magnetic
field model, twist breaks the symmetry with respect to the propagation direction
when the sign of one of the two wavenumbers is changed \citep[see also the
results of][for torsional Alfv\'en waves]{farahanietal2010}. The increase in
frequency of the modes $(m,k_1)$ and $(-m,-k_1)$ with respect to the un-twisted
case is around $4\%$ while the decrease for $(-m,k_1)$ and $(m,-k_1)$ is $8\%$
for $\alpha=0.01$. This means that a very weak twist (of about $1\%$) produces
an effect on the frequency that is not negligible. The maximum value of $\alpha$
in our calculations has been chosen to satisfy that $\alpha<\alpha_J$ and
$\alpha<\alpha_M$ for the reasons explained in
Section~\ref{secteigenmodestwist}.

Figure~\ref{omegalamb} also suggests that a linear variation of $\omega$ with
$\alpha$ is a good approximation of the actual dependency,
\begin{eqnarray} \omega &\simeq&k_1 c_{\rm k} 
+a\,\alpha,\,
{\rm for}\,\, (m,k_1),\, (-m,-k_1), \label{wp} \\ \omega &\simeq&k_1 c_{\rm k} 
-a\,\alpha, \,
{\rm for}\,\, (m,-k_1),\, (-m,k_1), \label{wm} \end{eqnarray}
$a$ being the slope
of the linear approximation.  The twist profiles considered by
\citet{goossensetal92} and \citet{ruderman07} are special since for those cases
$a=0$, i.e., there is no frequency change for different wavenumbers (for
example, $m=\pm 1$). However, for nonconstant twist configurations, like
those studied here, there is always a frequency split.

Similar behaviour occurs for different longitudinal wave numbers, the curves are
just shifted vertically (see continuous and dashed thin lines). The slope of the
curves is independent of the value of $k$, as we can see in
Figure~\ref{omegalamb} (compare the curves for $k_1$ and $k_2$), but as we  show
below, it depends on the particular twist profile. According to
Figure~\ref{omegalamb} for a fixed frequency and fixed value of twist, there are
always four possible solutions, $(m,k_1)$, $(-m,-k_1)$, $(-m,k_2)$, $(m,-k_2)$
(see the intersection of the solid thick and dashed thin lines around
$\omega=7.85\times 10^{-2}$ and $\alpha=3.5\times 10^{-3}$). This property will
be used as the basis to construct approximate standing solutions in the
following sections.

\begin{figure}[!ht] \center{\includegraphics[width=8.5cm]{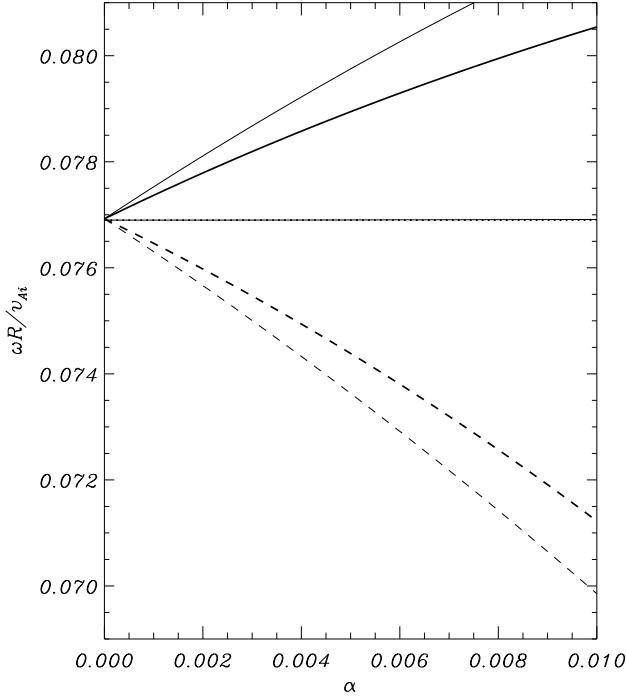}} 
\caption{\small Frequency as a function of twist ($\alpha$). For this particular
example, $\rho_{\rm i}/\rho_{\rm e}=3$, $p=1/2$, $q=3/2$ represented with thick
lines while the case $p=0$, $q=2$ is plotted with thin lines. The dotted line
represents the solution for the untwisted case. The horizontal continuous line is
obtained by numerically solving the eigenvalue problem in 2D. The same notation
as in Figure~\ref{omegalamb} is used.}\label{omega2D} \end{figure}

A qualitatively similar behaviour occurs when the parameters are changed. In
Figure~\ref{omega2D} the results are plotted for $p=0$ and $q=2$ (see thin
lines). Now the slope of the curves is steeper than for the case $p=1/2$ and
$q=3/2$ but the curves still show a linear behaviour with $\alpha$. Since for
the case $p=0$ and $q=2$ the region where the twist is localised is larger, the
effect on the frequency is more pronounced. Indeed, we tested different
nonconstant twist profiles and the curves are always linear with $\alpha$ or
equivalently with $B_\varphi$ (for $|B_{\varphi}|<<|B_z|$).

\begin{figure}[!ht] \center{\includegraphics[width=9cm]{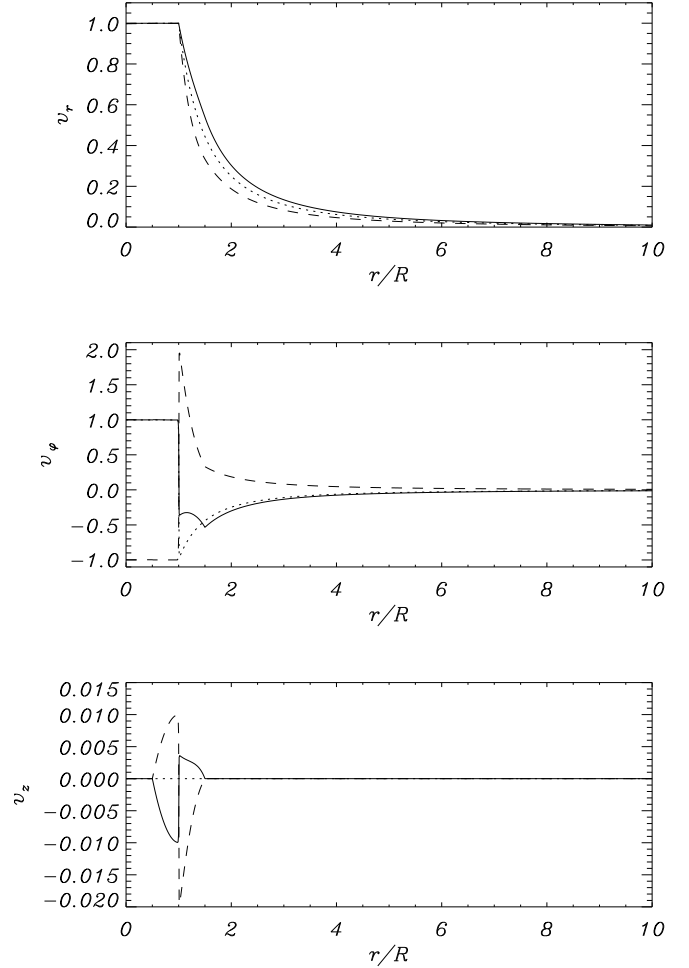}} 
\caption{\small Dependence of the three velocity components of the eigenfunction
with the radial coordinate. For this particular example, $\rho_{\rm i}/\rho_{\rm
e}=3$, $p=1/2$, $q=3/2$ and $\alpha=0.01$. 
The continuous and dashed lines correspond to
the modes $(m,k$) and $(-m,k)$, while the dotted line represents
the mode $(m,k$) in the absence of twist ($\alpha=0$). 
}\label{eigenfun} \end{figure}

We now aim to know the effect of twist on the eigenfunctions. It can be shown
from the MHD equations (see Appendix) that if  Fourier analysis is performed in
both the azimuthal and longitudinal direction, like in the present case, then
there is a phase shift between the radial component (pure real function) and the
azimuthal and longitudinal components (purely imaginary functions). In
Figure~\ref{eigenfun} the radial dependence of $v_r$, $v_\varphi$ and $v_z$ is
plotted for the modes $(m,k)$ (continuous line) and $(-m,k)$ (dashed line). The
same figure is obtained for the modes $(-m,-k)$ and $(m,-k)$. We see that the
radial dependence of the modes is essentially the same inside the tube and only
in the external medium near the tube boundary the dependence is slightly
different (compare also with the dotted line, which represents the eigenfunction
for the classical untwisted tube). The azimuthal velocity component shows more
relevant differences for the two modes. The mode  $(m,k)$ has a reduced shear at
the tube boundary while the opposite behaviour is displayed by the mode
$(-m,k)$, showing an increased shear with respect to the un-twisted case. This
behaviour is also present in the $v_z$ component, which is introduced by the
presence of twist. Since we are in the zero-$\beta$ approximation, the velocity
along the magnetic field is zero, i.e.,  $v_\parallel ={\bf {v}} \cdot {\bf
{B_0}}/{|B_0|}=0$, as expected. The $v_z$ component has a jump at the tube
boundary and is different from zero only in the region where there is twist (in
this particular example between $r=1/2R$ and $r=3/2R$). Notice that the
amplitude of this velocity component is rather small in comparison with $v_r$
and $v_\varphi$ since we are considering  a weak twist.

\section{Results: standing waves}

So far we focused on both azimuthally and longitudinally propagating waves.
Our interest is now on the analysis of standing waves, if they exist in the
twisted configuration.

\subsection{Standing in $\varphi$, propagating in $z$}

An azimuthally propagating  wave, for example the mode $m=1$, produces a motion
that displaces the whole tube and its axis is moving following a circular path
because of the propagating nature of the wave in $\varphi$. If the motion is
clockwise for $m$ then it is anti-clockwise for $-m$. For the tube to oscillate
transversally, the propagating $m$  mode  has to be combined with the $-m$ mode.
When there is no twist, the frequency of the mode $(m,k)$ is the same as the for
the mode $(-m,k)$ ($\omega=k c_{\rm k}$ in the thin tube limit), and the
superposition is trivial. In the presence of twist the situation is more
complicated since, as we have shown in Section~\ref{sectprop}, the modes $(m,k)$
and $(-m,k)$ have different frequencies. This means that with a single
longitudinal wavenumber it is not possible to have a standing solution in the
azimuthal direction. However, we have shown that we can always find another pair
of wavenumbers with the same frequency. This means that the linear combination
of the modes $(m,k_1)$ and $(-m,k_2)$ will lead to the standing solution in
$\varphi$ we are looking for. Formally, we have to combine the two waves, 
\begin{eqnarray}  y=f(r)\exp{i\left(\omega t +m\varphi+k_1
z\right)}+g(r)\exp{i\left(\omega t -m\varphi+k_2 z\right)},\label{standthet0}
\end{eqnarray} 

\noindent where $y$ is, for example, any of the velocity components of the
eigenfunction. Note that $f(r)$ and $g(r)$ are different functions since they
are associated to different pairs of wavenumbers $(m,k_1)$ and $(-m,k_2)$.
However, in the thin tube approximation, when $k_1 R \ll 1$ and $k_2 R \ll 1$,
the two functions are essentially the same ($f(r)\approx g(r)$ or $f(r)\approx
-g(r)$ as Figure~\ref{eigenfun} suggests). This is a nontrivial assumption,
which might not be a good approximation when we relax the thin tube and the
small $\alpha$ approximations.  Using this approximation, Eq.~(\ref{standthet0})
reduces to 
\begin{eqnarray} y=f(r)\cos\left(m\varphi + k_{\rm T} z\right)\exp{i\left(\omega t
+k z\right)}, \label{standthet} \end{eqnarray}
where we have introduced for convenience the following notation 
\begin{eqnarray}
k&=&\frac{k_1+k_2}{2},\label{kdef}\\
k_{\rm T}&=&\frac{k_1-k_2}{2}.\label{ktdef}
\end{eqnarray}
 If instead of the sum of the modes $(m,k_1)$ and $(-m,k_2)$ we choose the
difference as the linear combination, we find 
\begin{eqnarray} y=f(r)\sin\left(m\varphi + k_{\rm T}z\right)\exp{i\left(\omega t
+k z\right)}. \label{standthetsin} \end{eqnarray} 
It is straightforward to construct the standing solution in $\varphi$ but propagating in the
positive $z$-direction (note that Eq.~(\ref{standthet}) represents a wave
propagating in the negative $z$-direction). The solution is based on the superposition of the modes
$(-m,-k_1)$ and $(m,-k_2)$, and it is 
\begin{eqnarray} y=f(r)\cos\left(m\varphi + k_{\rm T}z\right)\exp{i\left(\omega t
-k z\right)}. \label{standthetappm} \end{eqnarray}
Again the sinus solution is obtained when we subtract the modes.

Equations~(\ref{standthet}) and (\ref{standthetappm}) represent propagating
waves in the $z-$direction and produce a transverse displacement of the tube
so that now it depends on $z$ through the term $k_{\rm T} z$. Thus, the direction of
the transverse oscillation of the loop varies as the wave propagates along the
tube. This represents a change in the polarisation of the transverse motion and
is one of the main effects of twist on the eigenmodes. For the untwisted tube
the polarisation of oscillation is always the same since $k_{\rm T}=0$.

The choice of $k_1$ and $k_2$ is not arbitrary since the modes with the pairs of
wavenumbers $(m, k_1)$ and $(-m, k_2)$ (the same applies to the combination
$(-m, -k_1)$ and $(m, -k_2)$) need to have the same frequency. The appropriate
wavenumbers can always be determined numerically for a given frequency, as
Figure~\ref{omegalamb} suggests. However, if $\alpha$ is small, this selection is
straightforward because we can use the approximations for the frequency (see
Eqs.~(\ref{wp})-(\ref{wm})) to calculate $k_1$ and $k_2$,
\begin{eqnarray} 
k_1&\simeq&\frac{\omega-a\,\alpha}{c_{\rm k}},\\
k_2&\simeq&\frac{\omega+a\,\alpha}{c_{\rm k}},
\end{eqnarray}
where $a$ is the slope of the curves. Now using Eqs.~(\ref{kdef}) and (\ref{ktdef}) we have  
\begin{eqnarray}\label{omegaapprox}
k&\simeq&\frac{\omega}{c_{\rm k}},\\
k_{\rm T}&\simeq&-\frac{a\,\alpha}{c_{\rm k}}\label{ktapprox}.
\end{eqnarray}
 
\noindent From the first equation we conclude that the dispersion relation of
the wave in the twisted case, $\omega=k c_{\rm k}$, is exactly the same as in
the untwisted tube (under the thin-tube approximation and for weak twist). Thus,
for weak twist, there is no effect on the frequency of the propagating waves in
$z$ and standing waves in $\varphi$. Nevertheless, from the second equation we
find that the polarisation of the motion shows a dependence with twist ($k_{\rm
T}\neq 0$). This dependence is linear with $\alpha$. For the un-twisted case
($\alpha=0$) we recover the well-known solution of the transverse kink mode
without variation of polarisation along the tube.

\subsection{Standing in $z$, propagating in $\varphi$}

Now we seek solutions that are standing in the $z-$direction. These
solutions have to satisfy line-tying conditions, i.e., the three velocity
components must be zero at the footpoints of the loop, located at $z=0$ and
$z=L$, where $L$ is the total length of the tube. We proceed as in the previous
section and select the appropriate superposition of modes. A suitable choice
is $(m,k_1)$ and $(m,-k_2)$,
\begin{eqnarray} y=f(r)\left(\exp{i\left(\omega t +m\varphi+k_1 z\right)}+\exp{i\left(\omega t
+m\varphi-k_2 z\right)}\right),\label{standthets} \end{eqnarray}
which can be written as
\begin{eqnarray} y=f(r)\cos\left(
k z\right)\exp{i\left(\omega t +m\varphi+k_{\rm T} z\right)}.
\label{standzp}
\end{eqnarray}
The other choice corresponds to the modes $(-m,-k_1)$ and $(-m,+k_2)$, which  
represent a propagating wave in the positive $\varphi-$direction,
\begin{eqnarray} y=f(r)\cos\left(
k z\right)\exp{i\left(\omega t -m\varphi-k_{\rm T} z\right)}.
\label{standzm}
\end{eqnarray}

\noindent As in the previous expressions, a sinus is obtained instead of a
cosinus when the difference of the modes is considered instead of the sum, 
\begin{eqnarray} y=f(r)\sin\left(
k z\right)\exp{i\left(\omega t +m\varphi+k_{\rm T} z\right)},
\label{standzpsin}
\end{eqnarray}
and
\begin{eqnarray} y=f(r)\sin\left(
k z\right)\exp{i\left(\omega t -m\varphi-k_{\rm T} z\right)}.
\label{standzmsin}
\end{eqnarray}
Strictly speaking these modes cannot be classified as fully standing modes,
because of the $z$ dependence in the exponential.

Applying line-tying conditions imposes restrictions on the
wavenumbers. Using the sinus solution given by Eq.~(\ref{standzpsin}) or
(\ref{standzmsin}) allows us to easily impose that the velocity must be zero
$z=0$ and $z=L$. The first condition is trivially satisfied while the second
leads to \begin{eqnarray} \sin\left( k L\right)=0, \end{eqnarray} and thus
\begin{eqnarray}  k=\frac{\pi}{L} n , n=1,2,3\dots.\label{linetying}
\end{eqnarray} \noindent This is essentially the same situation as in the
untwisted tube, we have a discrete set of wavenumbers satisfying
the boundary conditions. 

In constructing the standing solution in $z$ we have assumed that this solution
is the superposition of two propagating waves with the same frequency, different
longitudinal wavenumbers and the same radial dependence. As we have explained,
this is only true within limit $kR\ll 1$. In principle, we cannot assume a
Fourier analysis in the $z-$direction because of line-tying conditions and
consider single modes if we aim to obtain the full solution to the problem. For
this reason, we now calculate the eigenmodes by solving the problem in $r$ and
$z$ for a fixed $m$. This will allow us to compare the results with the
semi-analytical results given by Eqs.~(\ref{omegaapprox}) and (\ref{ktapprox}).
We used the two-dimensional version of the code PDE2D to calculate the modes
with line-tying conditions at $z=0$ and $z=L$. In Figure~\ref{omega2D} the
eigenfrequency is plotted as a function of twist. Clearly, the frequency is
almost independent of $\alpha$ (see the continuous horizontal line), which 
agrees with the analytical results obtained within the limit of small $\alpha$ 
given by Eq.~(\ref{omegaapprox}) (compare also the continuous horizontal line
with the dotted line, which corresponds to the untwisted case). Therefore, it is
a good approximation (within the limit of small $\alpha$) to assume that we can
construct standing eigenmodes by summing two propagating waves with the same
frequency but slightly different wavenumbers.

\begin{figure}[!ht] \center{\includegraphics[width=8cm]{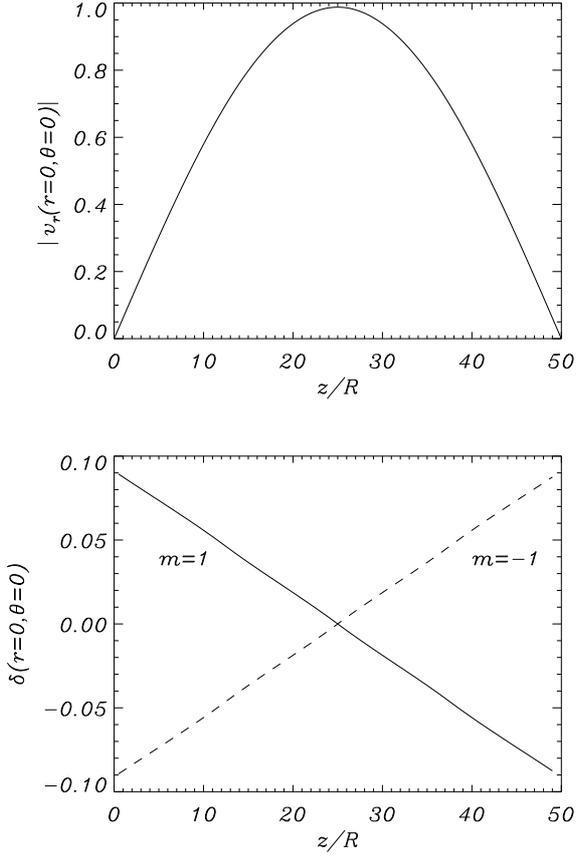}} 
\caption{\small Eigenfunction as a function of $z$ at the loop axis calculated
numerically using line-tying conditions. The modulus and phase are plotted for
the modes $m=\pm 1$.}\label{eigenfunctcut} \end{figure}

A plot of the eigenfunction at $r=0$ and $\varphi=0$ as a function of $z$ is
shown in Figure~\ref{eigenfunctcut}. It is worth to mention that for the 2D
problem the eigenfunction is always a complex number (this can be seen from
Eqs.~(\ref{eqvrapp})-(\ref{eqbzapp}) in the appendix) although the frequency is
still a real magnitude since we do not consider unstable modes or resonances.
For this reason we represent the modulus and the phase ($\delta$) of the
eigenfunction in Figure~\ref{eigenfunctcut}. The modulus has a sinusoidal
profile while the phase shows a linear dependence with $z$. This is exactly the
behaviour predicted by Eq.~(\ref{standzpsin}). The slope of the phase is
precisely the magnitude $k_{\rm T}$, which is negative according to
Eq.~(\ref{ktapprox}). The $m=-1$ mode shows the expected positive slope, in
agreement with Eqs.~(\ref{standzmsin}) and (\ref{ktapprox}).

\subsection{Standing in $\varphi$ and $z$} 

Once we have derived the solutions that are standing, or equivalent to
standing, in one of the directions, it is
straightforward to address the full standing problem. There are two ways to
proceed that must be equivalent, either we add the solutions that are
standing in $\varphi$ but propagating in $z$, or we combine the solutions that are
(almost) standing in $z$ but propagating in $\varphi$. For example, from the combination
of the solution given by Eq.~(\ref{standthet}), representing a standing wave in
$\varphi$ but propagating in $z$,  minus the complementary wave travelling in the
opposite $z-$direction given by Eq.~(\ref{standthetappm}), we find the ``full'' standing solution
\begin{eqnarray} y=f(r)\cos\left(m\varphi +
k_{\rm T} z\right)\sin\left(
k z\right)\exp{\left(i \omega t\right)}.
\label{standthetsinm}
\end{eqnarray}
It is easy to check that we obtain exactly the same expression by combining the
solutions given in Eqs.~(\ref{standzpsin})-(\ref{standzmsin}) for mainly
representing standing waves in $z$ but propagating in $\varphi$.

Based on Eq.~(\ref{standthetsinm}) and using the proper phase shifts between the
components of the displacement, we obtain the following eigenfunction
\begin{eqnarray}
\xi_r&=&\tilde f(r)\cos\left(m\varphi +
k_{\rm T} z\right)\sin\left(k z\right)\sin \left(\omega t\right),\label{xir}\\
\xi_\varphi&=&\tilde g(r)\sin\left(m\varphi +
k_{\rm T} z\right)\sin\left(
k z\right)\sin \left(\omega t\right),\label{xivarphi}\\
\xi_z&=&\tilde h(r)\sin\left(m\varphi +
k_{\rm T} z\right)\sin\left(
k z\right)\sin \left(\omega t\right).\label{xiz}
\end{eqnarray}
Similar expressions have been used in the past as approximate solutions to test the stability of different 
twisted magnetic configurations. However, as far as we know,
these expressions have not been applied to stable kink waves.

From Eqs.~(\ref{xir})-(\ref{xiz}) we can easily calculate the displacement of
the tube axis, which is a relevant quantity when it comes to observations of transverse loop waves. According
to Figure~\ref{eigenfun},  inside the tube $\tilde f(r)\simeq
\tilde g(r)=const$ in the thin-tube approximation, and the displacement in $z$
has a node at the axis. Thus, in Cartesian coordinates at the
axis ($x=y=0$) \begin{eqnarray}\xi_x&=&\xi_0 \cos\left(k_{\rm T}
z+\varphi_0\right) \sin\left(k z\right)  \sin\left(\omega t\right),\\
\xi_y&=&\xi_0  \sin\left(k_{\rm T} z+\varphi_0\right) \sin\left(k z\right)
\sin\left(\omega t\right),\\ \xi_z&=&0, \end{eqnarray}  $\xi_0$ being an
arbitrary amplitude and $\varphi_0$ an arbitrary  phase. These expressions
show that the tube does not move along the axis while the displacement in the
plane perpendicular to the axis is given by the product of two sinusoidal
terms, one involving a scale associated to the twist, $k_{\rm T}$, and another
related to the longitudinal wavenumber, $k$. At the axis
\begin{eqnarray}
\sqrt{\xi_x^2+\xi_y^2+\xi_z^2}=\xi_0 \sin\left(k z\right) 
\sin\left(\omega t\right),
\end{eqnarray}
meaning that the total displacement is exactly the same as in the un-twisted
case. The difference is in the components of the displacement which have a
different weight depending on the position along the tube.

\begin{figure}[!ht] \center{\includegraphics[width=9cm]{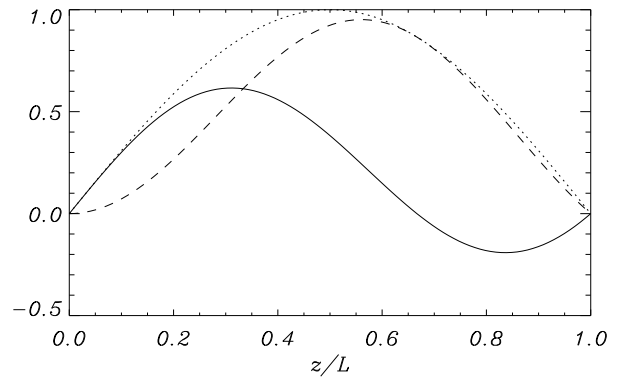}} 
\caption{\small Displacement of the tube axis as a function of position along
the twisted tube. The continuous line corresponds to the displacement in the
$x-$direction while the dashed line represents the displacement in the
$y-$direction. The dotted line is the purely sinusoidal displacement in the $x-$
direction for the untwisted case. In this example, $k_{\rm T}=-0.75k$,
$\varphi_0=0$, where
$\xi_0/L=1$.} \label{eigen0_5} \end{figure}

\begin{figure}[!ht]
\center{\includegraphics[width=9cm]{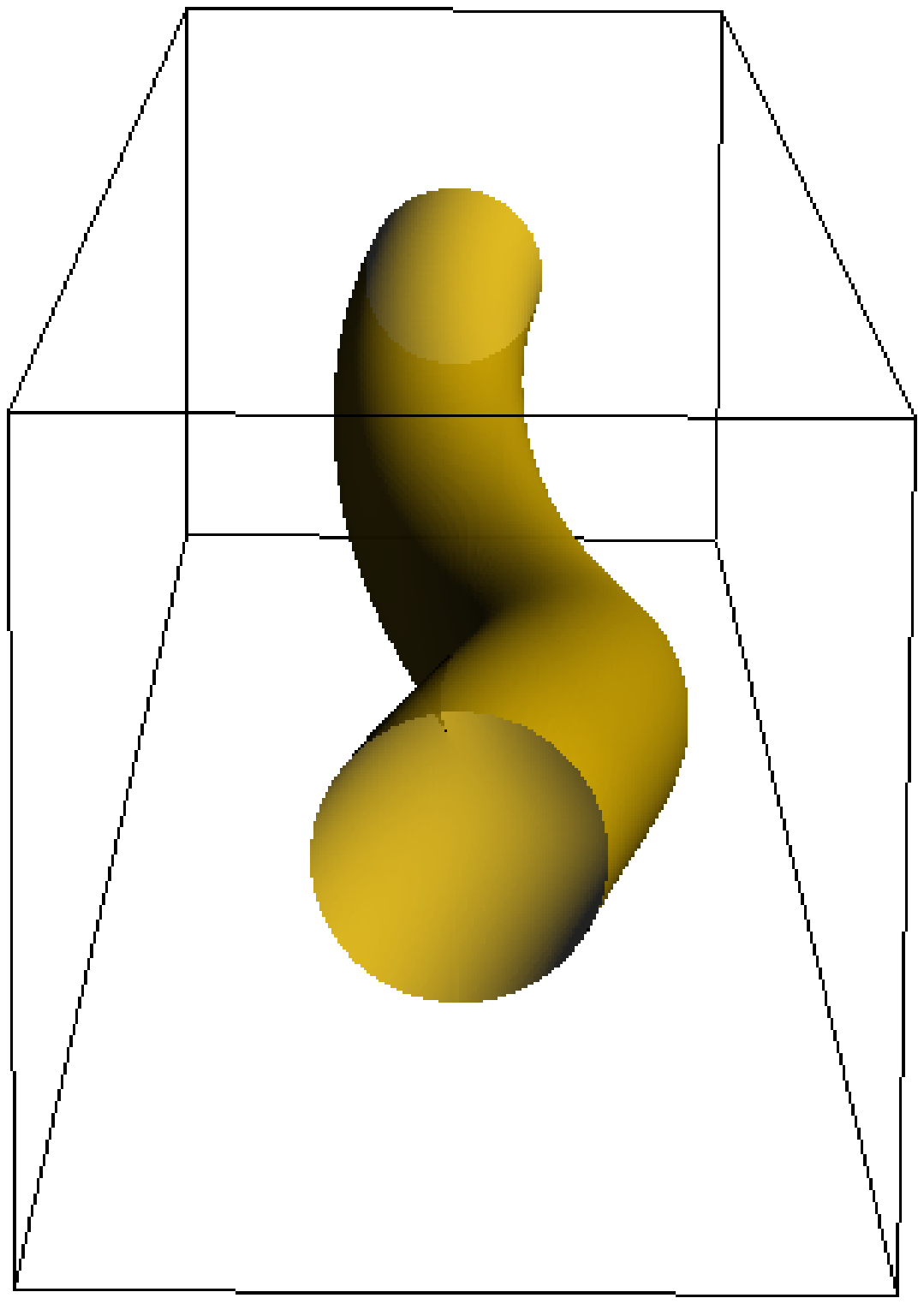}}
\center{\includegraphics[width=9cm]{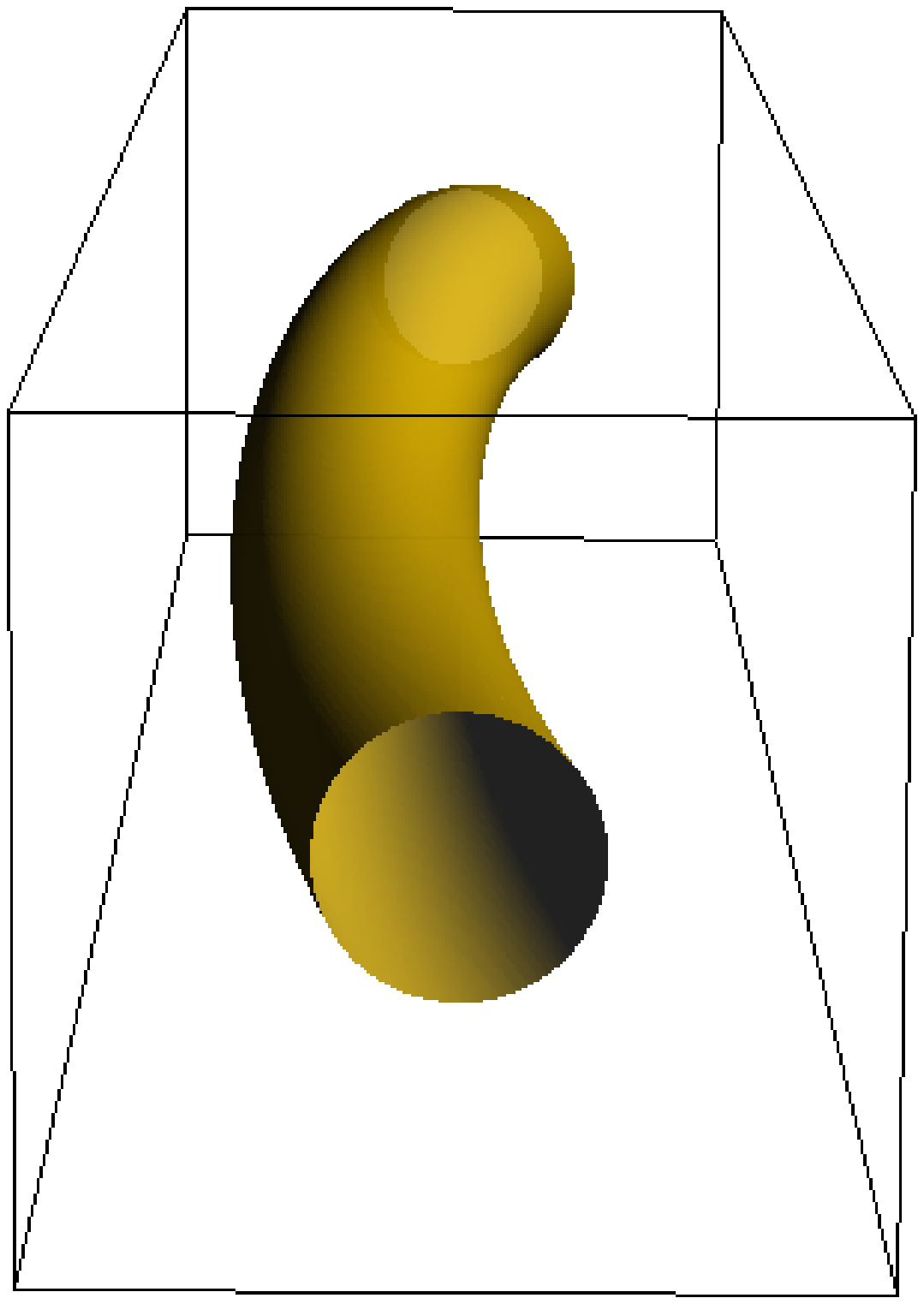}} 
\caption{\small Snapshots of the displacement of the tube with twisted magnetic
field at two different
times during one oscillation period. In this example, $k_{\rm T}=-0.75k$. The $x-$axis is horizonal, the
$y-$axis is vertical and the $z$-axis is perpendicular to the $xy$-plane. Line-tying
conditions are applied at the front and back planes ($z=0$ and $z=L$).} \label{eigen0_75}
\end{figure}

In Figure~\ref{eigen0_5} the displacement of the tube axis is plotted for a
situation with moderate twist. Although the calculations in the previous
sections were made very weak twist, we expect that for higher values of the
$B_\varphi$ component the motion of the axis is qualitatively the same. The
differences in the polarisation of the motion with respect to the untwisted
situation (see dotted line) are evident in this plot. The motion of the tube is
even more clear in Figure~\ref{eigen0_75} where the displacement of the tube is
represented in three-dimensions at two different times during the oscillation.
We clearly see the departures from the purely sinusoidal displacement for the
untwisted case. In the top panel of Figure~\ref{eigen0_75} the displacement is
closer to the first longitudinal harmonic than to the fundamental mode. However,
at later stages of the oscillation, see bottom panel of Figure~\ref{eigen0_75},
the motion resembles the fundamental mode. Thus, twist can have a pronounced
effect on the displacement of the tube axis during the oscillation.  

\section{Conclusions and discussion}

From inspection of the local Alfv\'en frequency it might be anticipated that an
azimuthal magnetic field can have a relevant effect on MHD waves even if
$|B_\varphi|<<|B_z|$. It is the competition between the two quantities $k R$ and
$\alpha=B_{\varphi,max}/B_z(0)$ that determines the extent of the effect. In
this work the quantity $\alpha$ is a small quantity by choice since we aim to
focus on magnetic fields that are dominated by the longitudinal components. The
quantity $kR$ is often small as most of the loops are much longer than wide or
the longitudinal wavelength is much longer that the radius of the loop. So it is
really a competition between two small quantities $kR$ and $\alpha$. In the
present investigation we have deliberately excluded the regime $\alpha/kR\sim 1$
and focused our attention on $\alpha<\alpha_J$ and $\alpha<\alpha_M$. In this
way we avoid the interesting complications of resonant absorption. However,
restricting ourselves to these values of $\alpha$ means that the azimuthal magnetic fields
are in a sense very weak, and that the effects on the MHD waves by azimuthal
magnetic field are less pronounced than they effectively are.

In general, the frequencies of MHD waves in models with nonconstant magnetic
twist are nondegenerate compared to their counterparts in models with constant
magnetic twist.  This means that the frequencies of propagating waves with
wavenumbers $(m,k)$ and $(-m,-k)$ are equal but different from the frequency of
the modes $(-m,k)$ and $(m,-k)$. Twist breaks the symmetry with respect to the
propagation direction. This property has consequences for standing modes that
require the superposition of propagating waves. We have shown, using an
approximate method valid for weak twist, that the frequency of standing waves
is unaffected by the presence of twist. Under such conditions the tube
oscillates transversally at the characteristic kink frequency.


However, the most interesting effect of magnetic twist on standing oscillations
is in the change of polarisation of the transverse motion along the tube. We
have analytically demonstrated that the change in the direction of the
transverse displacement of the tube is linearly proportional to the twist.
Moreover, the polarisation of the displacement of the axis along the tube is
simply described by the product of two sinusoidal terms involving a wavenumber
associated to the twist, and a longitudinal wavenumber that satisfies the
line-tying boundary conditions at the footpoints of the loop \citep[this is very
similar to the situation for a tube with a siphon flow along the axis studied
by][]{terradasetal2011}. For a very weak twist, for instance 1/10 windings of
the field along the loop axis, twist apparently introduces a motion
perpendicular to the dominant direction of oscillation that is around nine times
smaller in amplitude. The effect on the polarisation of the motion is small
because the twist in the tube is very weak in this particular example. When the
twist increases, the change in polarisation becomes stronger, as is shown in
Figures~\ref{eigen0_5} and~\ref{eigen0_75}. An important result here is that a
weak twist can produce displacements in {\em any direction} perpendicular to the
unperturbed tube axis. Thus, in contrast to the effect of curvature of the 
magnetic field lines and the effect of nonuniform cross section, which both
introduce vertical and horizontal modes, twist produces displacements in all
directions. This agrees with the results of \citet{ruderscott2011}, who
considered kink oscillation of a nonplanar coronal loop involving magnetic
twist. In view of these results the role of twist on stable transverse motions
has been underestimated in the context of loop oscillations. Potentially, the
signatures of twist on standing transverse oscillations could be used as a way
to infer the value of the azimuthal component of the magnetic field in coronal
loops. This could have clear seismological applications. However, the fact that
real coronal loops are in many cases nonplanar and noncircular, i.e., they have
an helical shape \citep[see the review of][]{asch2009}, might complicate the
comparison between theory and observations.

Finally, it is worth to notice that the analysis performed in this paper is
based on the assumption of very weak twist. This has allowed us to avoid some
complications that appear when the $B_\varphi$ component of the magnetic field
is increased. First of all, Alfv\'enic resonances cannot be avoided for
moderate twist \citep[see the work of][]{karamibaha10}. Second, the modes
might become kink-unstable. A detailed investigation of these effects on
standing kink modes need to be addressed in future works.

{\acknowledgements J.T. acknowledges support from the Spanish Ministerio de
Educaci\'on y Ciencia through a Ram\'on y Cajal grant. Funding provided under
the project AYA2011-22846 by the Spanish MICINN and FEDER Funds, and the
financial support from CAIB through the ``Grups Competitius'' scheme and FEDER
Funds is also acknowledged by J.T. M.G. acknowledges financial support received
during his visits at Universitat de les Illes Balears (UIB), funding under grant
AYA2011-22846 and University of Leuven GOA 2009/009 is also acknowledged. The
authors also thank J. Andries for his comments and suggestions during the
preparation of this paper.}

\appendix
\section{}
The ideal linearised MHD equations in the zero-$\beta$ regime in the presence
of magnetic twist are the following:
\begin{eqnarray}
i \omega v_r &=&\frac{1}{\mu_0 \rho}
\left(\left(\frac{\partial b_r}{\partial{z}}-\frac{\partial b_z}{\partial{r}}\right) B_{z}
-\frac{dB_{z}}{dr}b_z\right.\nonumber \label{eqvrapp}\\ 
& & \left. -\frac{1}{r}\left(\frac{\partial
r b_\varphi}{\partial{r}}-i m b_r\right)B_{\varphi} -\frac{1}{r}\frac{d(
r B_{\varphi})}{dr} b_{\varphi}\right),\label{vreqn}\\
i \omega v_\varphi&=&\frac{1}{\mu_0 \rho}\left(-\left(\frac{i m}{r}b_z
-\frac{\partial b_\varphi}{\partial{z}} \right)B_{z}+\frac{1}{r}\frac{d(r
B_{\varphi})}{dr} b_{r}\right),\\
i \omega v_z&=&\frac{1}{\mu_0 \rho}\left(\left(\frac{i m}{r}b_z-
\frac{\partial b_\varphi}{\partial z} \right)B_{\varphi}+\frac{dB_{z}}{dr} b_{r}\right),\\
i \omega b_r&=&B_{z} \frac{\partial v_r}{\partial z}+B_{\varphi} \frac{i m}{r} v_r,\\
i \omega b_\varphi&=&B_{z} \frac{\partial v_\varphi}{\partial z}-B_{\varphi}
\frac{\partial v_z}{\partial z}-B_{\varphi}\frac{\partial
v_r}{\partial{r}}-v_r\frac{dB_{\varphi}}{dr},\\
i \omega b_z&=&\frac{1}{r}\left(-B_{z} \frac{\partial
r v_r}{\partial{r}}- r v_r \frac{dB_{z}}{dr}-B_{z} i m v_\varphi +B_{\varphi} i
m v_z\right),\label{eqbzapp}
\end{eqnarray}
\noindent where ${\bf v}=(v_r,v_\varphi,v_z)$ is the velocity and ${\bf
b}=(b_r,b_\varphi,b_z)$ is the perturbed magnetic field. If a Fourier analysis is
performed in the $z-$direction, the derivatives with respect to $z$ have to be
replaced by $i k$.

\bibliographystyle{aa}      
\bibliography{jaume}   

\end{document}